\documentclass{article} 
\usepackage{GEM_workshop_2026,times}

\usepackage{amsmath,amsfonts,bm}

\def\eqref#1{equation~\ref{#1}}

\def\1{\bm{1}}

\DeclareMathAlphabet{\mathsfit}{\encodingdefault}{\sfdefault}{m}{sl}
\SetMathAlphabet{\mathsfit}{bold}{\encodingdefault}{\sfdefault}{bx}{n}

\usepackage{hyperref}
\usepackage{url}
\usepackage{tabularx,booktabs}
\usepackage{graphicx}
\usepackage{lipsum}
\usepackage[utf8]{inputenc}
\usepackage{amsmath,amssymb}
\usepackage{bm}
\usepackage{algorithm}
\usepackage{algorithmic}
\usepackage{placeins}
\usepackage{dsfont}
\usepackage{url}
\usepackage{svg}
\usepackage{graphics}
\usepackage[export]{adjustbox}
\usepackage{soul}
\usepackage{bbding}
\usepackage{dsfont}
\usepackage[labelfont=bf]{caption}
\usepackage{wrapfig}
\usepackage{subcaption}
\usepackage{natbib}

\usepackage{booktabs} 
\usepackage{siunitx} 
\usepackage{amsmath} 

\newcommand\paulina[1]{{\textcolor{violet}{[PS #1]}}}

\newcommand{\ours}{FoldSAE}

\title{FoldSAE: Learning to Steer Protein Folding Through Sparse Representations}

\author{Wojciech Zarzecki \\
  University of Warsaw, \\
  Warsaw University of Technology
  \And
  Paulina Szymczak \thanks{Corresponding author. Email: \href{mailto:paulina.szymczak@helmholtz-munich.de}{paulina.szymczak@helmholtz-munich.de}} \\
  Helmholtz Munich \\
  \texttt{paulina.szymczak@helmholtz-munich.de} \\
  \And
  Ewa Szczurek \\
  University of Warsaw, \\
  Helmholtz Munich 
  \And
  Kamil Deja \\
  Warsaw University of Technology, \\
  IDEAS Research Institute 
}

\iclrfinalcopy

\begin{document}

\maketitle

\vspace{-1em}

\begin{abstract}

While models like RFdiffusion excel at generating protein backbones, their "black box" nature currently restricts design to a process of stochastic sampling rather than precise engineering. To bridge this gap, we introduce FoldSAE, a framework that adapts Sparse Autoencoders (SAEs) to decompose RFdiffusion’s dense activations into interpretable, monosemantic features. We demonstrate that these unsupervised features capture fundamental physical properties, including secondary structure formation and solvent-accessible surface area (SASA). Leveraging these insights, we implement a steering mechanism that enables targeted modulation of backbone folding and surface exposure during the denoising process, both in \textit{de novo} and anlogue generation. Our work pioneers a new framework for making RFdiffusion more interpretable, demonstrating how understanding internal features can be directly translated into precise control over the protein design process.

\end{abstract}

\vspace{-1em}

\section{Introduction}

The "black box" nature of deep learning methods presents a significant barrier in their adaptation in life science domains. While state-of-the-art models like RFdiffusion~\citep{watson2023de} show remarkable capabilities in generating novel protein backbones, our inability to understand their internal representations limits scientific insight and practical control. This lack of transparency, means we cannot debug, validate or steer the generative process itself, turning protein design into a matter of sampling and filtering rather then precise engineering.

Mechanistic interpretability aims to solve this issue, by finding human-understandable mechanisms within a model. A promising technique in this area is the Sparse Autoencoder (SAE)~\cite{Olshausen1997SparseCW}, which decomposes model's dense \textbf{activations} (the vector outputs of network layers during a forward pass) into a sparse set of "mono-semantic" \textbf{features} (directions in the activation space corresponding to distinct, interpretable concepts). This approach has provided unprecedented insight into language models~\citep{huben2024sparse,bricken2023monosemanticity,marks2025sparse}, with some applications in diffusion models~\citep{surkov2024unpacking,kim2024revelio,cywinski2025saeuron}.  In the molecular domain, however, existing work has been restricted to sequence-level representations~\citep{adams2025mechanistic,simon2025interplm,rives2021esm2}, leaving the potential for interpretable control over structure generation largely unexplored.

In this work, we introduce \ours{}, a new framework that leverages sparse autoencoders to interpret the protein folding process within RFdiffusion~\cite{watson2023de}. Our goal is to decompose RFdiffusion's complex, dense representations into a sparse set of monosemantic features, thereby uncovering its inner workings. To validate that these unsupervised features are indeed meaningful and useful, we train a single SAE that we use to analyse (1) the process of secondary structures generation and (2) solvent accessible surface area (SASA) as two complementary proof-of-concept applications. To that end, we first propose a simple heuristics based on block-ablation, to localize the specific parts of the model that are critical for those tasks. We use activations from the most relevant block to train a SAE, and use simple linear probing models to identify which of the discovered sparse features correlate with the target outcomes. Our analysis reveals that, even though trained in a fully unsupervised way, the same features often control both helix and strand formation, but with opposite correlation. Crucially, because the SAE is trained without supervision, the same model can be repurposed to identify features predictive of SASA, enabling control over backbone exposure without retraining. This confirms that a single SAE trained on one block captures a rich set of structural properties.

Those observations allow us to demonstrate that interpretability can be directly translated into precise control. We introduce a steering mechanism, where we can amplify or suppress these specific features during the diffusion-denoising process. As a result, we can precisely modulate the final protein structure, for example, by selectively reinforcing the features positively correlated with helices and blocking those correlated negatively, we can increase the amount of helices in generated protein backbones. Similarly, selecting features based on their correlation with SASA allows us to modulate the degree of backbone exposure in the generated designs. \ours{} thus offers a novel framework that directly links internal model representations to precise control, enabling a more directed protein design process. To facilitate future research, we release our code together with weights of trained SAE models at \href{https://github.com/wz7475/SAEtoRuleRFDiffusion}{\color{blue}{GitHub}}.


Our contribution can be summarized as follows: (i) we introduce \ours{}, a method for training a Sparse Autoencoder using the internal activations of RFdiffusion, successfully decomposing its dense representations into sparse, interpretable features; (ii) we establish a link between specific internal features and protein secondary structure, as well as SASA; and (iii) we design a steering mechanism that allows for precise, tunable control over the secondary structure formation and backbone solvent exposure during the RFdiffusion generative process.

\section{Background \& Related work}

\vspace{-0.5em}
\subsection{Protein preliminaries}
\vspace{-0.5em}
\textbf{Protein structure} is hierarchically organized: the amino acid sequence dictates local secondary structure elements (helices, strands, and coils) whose spatial packing determines the overall three-dimensional fold and, ultimately, biological function. Computationally, this structure can be reduced to the protein backbone, a repeating chain of three atoms per residue: the amino-group nitrogen ($N$), the alpha-carbon ($C_{\alpha}$), and the carbonyl carbon ($C$). This representation abstracts away side-chain identity, treating all residues as glycine, with the $N\text{-}C_{\alpha}\text{-}C$ coordinates fully specifying each residue's position and orientation in 3D space.


    \textbf{RFdiffusion} generates protein backbones by iteratively denoising a rigid-frame representation consisting of $C_{\alpha}$ coordinates and rotation matrices for each residue. The architecture utilizes 36 stacked blocks that simultaneously refine 1D, 2D, and 3D inputs to predict the necessary translations and rotations for the denoising process. This process yields a backbone structure that requires subsequent processing by models like ProteinMPNN~\citep{dauparas2022robust} to assign specific amino acid identities.

\vspace{-1em}
\subsection{Mechanistic interpretability}    



\vspace{-0.5em}

Within mechanistic interpretability methods, Sparse Autoencoders (SAEs)~\citep{Olshausen1997SparseCW} have emerged as a powerful tool to disentangle dense representations, into steerable features. This approach, specifically with top-k sparsity~\cite{Makhzani2013kSparseA}, have been successfully applied to Large Language Models~\citep{huben2024sparse,bricken2023monosemanticity,marks2025sparse}, while recently, works by \citet{surkov2024unpacking}, \citet{kim2024revelio}, and \citet{cywinski2025saeuron} have also demonstrated that the same tool can be used for precise control in image diffusion models. In biology, the utility of SAE has not been fully discovered. While works like InterProt~\citep{adams2025mechanistic} and InterPLM~\citep{simon2025interplm} utilize SAEs on ESM2 embeddings~\citep{rives2021esm2} for sequence design, they leave structure generation unexplored~\citep{garcia2025interpreting}, a gap this work fills by introducing interpretable steering to RFdiffusion.

\vspace{-1em}

\section{Method}


\vspace{-1em}

Our method aims to find features within the model activations, which encode information about interpretable properties of protein backbones to use them for steering during the inference process. It consists of three stages: \textbf{localization}, \textbf{interpretation} and \textbf{intervention}. In the stage of localization, we identify the crucial block encoding semantic information about protein backbone design. In the interpretation stage, we decompose activations of the chosen block and select those corresponding to the interesting target. Finally, in the intervention stage, we manipulate the identified features to steer the generation towards desired properties.    

\vspace{-1em}

\subsection{Localization} \label{blocks_ablation}
    The first crucial design choice is the selection of the block to intervene. To that end, we adapt the method introduced by~\citet{cywinski2025saeuron}, and determine 
    which block encodes information about desired properties, by performing iterative ablation of the blocks. We understand ablation of $n^{th}$ block as substituting its output with an output of the previous - $n^{th-1}$ block. As shown by~\cite{cywinski2025saeuron}, if a given block adds information about desired properties, its ablation should result in observable distribution shift in the final generations. 
    Formally, let $S$ denote a metric quantifying the target property for a given model configuration.
    We identify the optimal block index $m^*$ by finding the ablation that maximizes the metric deviation: $m^* = \operatorname*{argmax}_{m} | S(\mathcal{M}_{\text{orig}}) - S(\mathcal{M}_{\setminus m}) |$, where $\mathcal{M}_{\text{orig}}$ denotes the original model and $\mathcal{M}_{\setminus m}$ the model with the $m$-th block ablated.

\subsection{Interpretation} \label{selection_of_features}
\begin{wrapfigure}{r}{0.6\linewidth}
    \vspace{-20pt} 
    \centering
    \includegraphics[width=\linewidth]{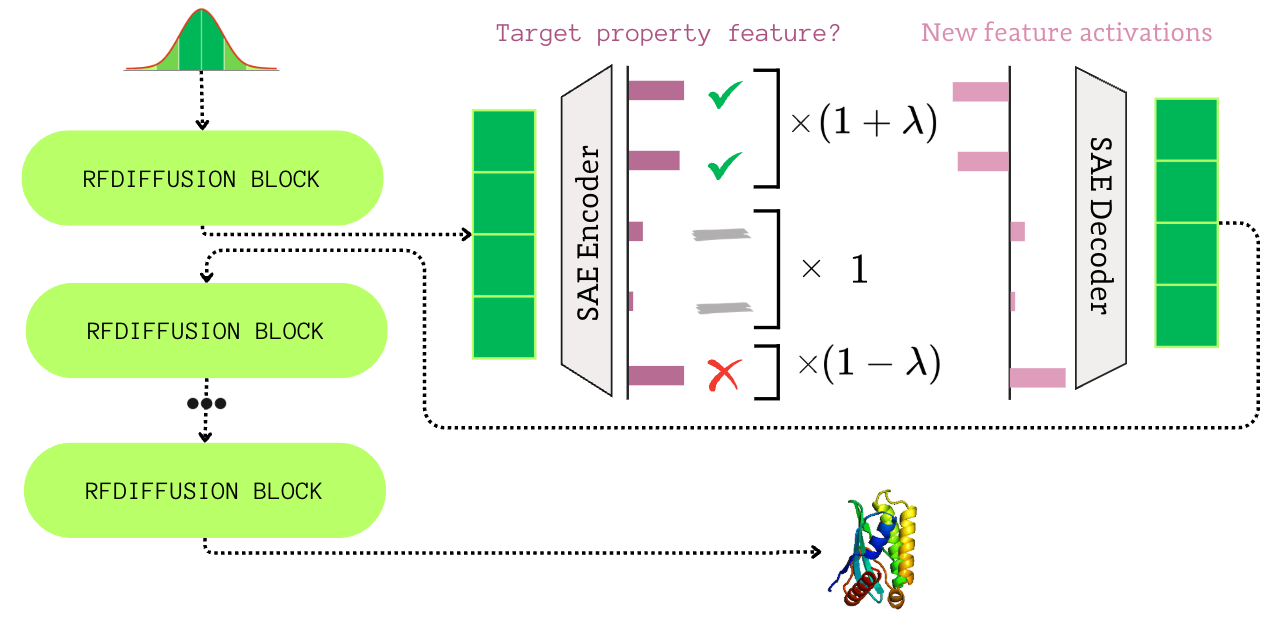}
    \caption{\textbf{Overview of the FoldSAE steering mechanism.} During the protein backbone generation process, activations from the localized RFdiffusion block are intercepted and decomposed into sparse features by the SAE Encoder. These features are then modulated based on their correlation with the desired target property (identified via probing classifiers). To steer the trajectory, features positively correlated with the target are amplified by a factor of $(1+\lambda)$, while negatively correlated features are suppressed by $(1-\lambda)$; neutral features remain unmodified (scaled by $1$). The adjusted features are reconstructed by the SAE Decoder and reintroduced into the network to guide subsequent diffusion steps.
    }
    \label{fig:sae_intervention}
    \vspace{-10pt} 
\end{wrapfigure}

    Given the activations from the selected block, our goal is to decompose them into interpretable features by training a SAE.
    
    \label{SAE_training_method}
    We train a Top-K SAE to reconstruct activations on a per-residue basis. We flatten the block outputs into $l$ sequential segments, treating each residue's embedding as an independent patch $\mathbf{x} \in \mathbb{R}^d$. The model consists of a single-layer encoder that encode activations $x$ into features: $\mathbf{z} = \text{TopK}(\text{ReLU}(\mathbf{W}_{\text{enc}}(\mathbf{x} - \mathbf{b})))$ and decoder that reconstructs them as $\mathbf{\hat{x}} = \mathbf{W}_{\text{dec}}\mathbf{z} + \mathbf{b}$, where $\mathbf{W}_{\text{enc}} \in \mathbb{R}^{n \times d}$ and $\mathbf{W}_{\text{dec}} \in \mathbb{R}^{d \times n}$ are weight matrices, and $\mathbf{b} \in \mathbb{R}^d$ is a learnable bias. The TopK operation enforces sparsity by zeroing out all but the $k$ highest latent values. The size of the latent dimension $n$ is a result of scaling $d$ by a fixed \textit{expansion factor}. We optimize the model by minimizing the reconstruction error objective: $\mathcal{L}(\mathbf{x}) = \|\mathbf{x} - \mathbf{\hat{x}}\|_2^2$. A well-trained SAE effectively decomposes dense activations into a dictionary of $n$ sparse, monosemantic feature vectors (columns of $\mathbf{W}_{\text{dec}}$), which can be directly manipulated to steer the generative process. 
    \vspace{-1em}

\subsection{Intervention}

    Let us assume, that we have identified a set of interesting SAE features that correlate (positively or negatively) with the desired property. We can use those features to steer the generation process through interventions by
    passing all of the activations through the autoencoder, while
    suppressing features negatively correlated with the target property, and reinforcing the positively correlated ones, leaving all others unchanged. We introduce a hyper-parameter, $\lambda$, that controls both the direction and magnitude of the intervention: $\lambda = 0$ corresponds to no modification, $\lambda > 0$ steers toward the target property, and $\lambda < 0$ steers away from it. Concretely, each SAE feature is scaled by $(1+\lambda)$ if positively correlated with the target, $(1-\lambda)$ if negatively correlated, and left unchanged otherwise (see Figure~\ref{fig:sae_intervention}).

    \vspace{-1em}

\section{Experiments}
    \vspace{-1em}
\subsection{Localization}
    To identify the specific RFdiffusion block responsible for encoding our target properties--secondary structure and SASA--we conduct a systematic ablation study as outlined in Section~\ref{blocks_ablation}. Secondary structure is annotated using STRIDE~\cite{frishman1995knowledge} with the standard eight-to-three-state reduction~\cite{rost1993prediction}, while SASA is computed via the Shrake--Rupley algorithm as implemented in FreeSASA~\cite{mitternacht2016freesasa} (see Appendix~\ref{app:labels}).
    By iteratively removing blocks and evaluating the generated structures, we observe that the ablation of block $\textit{main\_04}$ yields the most profound impact.
    Specifically, it renders the model incapable of generating alpha-helices and induces the largest deviation in SASA distributions compared to the baseline (see Appendix, Figure~\ref{fig:ablation_results} and Figure~\ref{fig:ablation_results_sasa}).
    Consequently, we select block $\textit{main\_04}$ for all subsequent interpretation and intervention experiments.

\subsection{Interpretation}

    \paragraph{Feature selection}

    Using the single SAE trained in the previous step, we identify features that discriminate between classes of our target properties by fitting logistic regression models to the fixed SAE's latent features to predict the corresponding property class. We describe details of dataset gathering and models training in Appendix~\ref{probing_detials}.
    In our experiments, we identify features that discriminate between binary target classes. For secondary structures, we utilize 'helix vs. others' and 'strand vs. others' classifiers; for SASA, we establish low and high categories based on the 25th and 75th percentiles of observed values.
    We select indices where both models' coefficients exceed the threshold and have \textbf{opposite signs}.
    For example, a feature that is strongly positive for the helix classifier and strongly negative for the strand classifier is highly discriminative between them.
\vspace{-1em}

\subsection{Intervention}

    \begin{wrapfigure}{r}{0.6\linewidth}
        \centering
        \vspace{-10pt} 
        \includegraphics[width=1\linewidth]{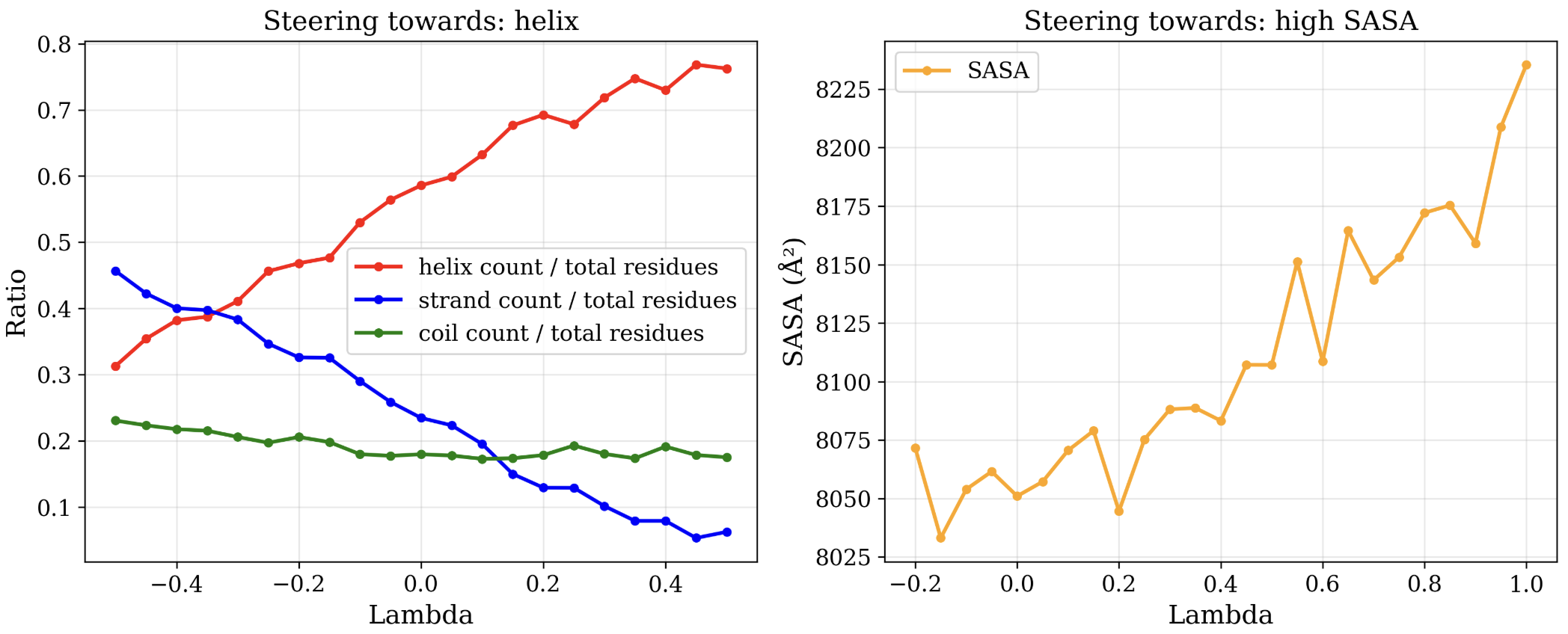}
        \caption{\textbf{FoldSAE intervention results.} (Left) Fraction of residues assigned to helices (red), strands (blue), and coils (green) as a function of steering intensity $\lambda$ when steering towards helices. (Right) Solvent Accessible Surface Area (SASA) of generated backbones as a function of $\lambda$ when steering towards high-SASA features, showing a clear ability to modulate surface exposure.}
        \label{fig:intervention_combined}
        \vspace{-10pt} 
    \end{wrapfigure}

    To evaluate whether the discovered features are causxnally linked to generation outcomes, we perform a series of interventions to steer the process toward specific secondary structures or surface properties. We vary the steering intensity $\lambda$ across empirically determined ranges. To ensure precision, prior to intervention, we employ pre-trained linear regressors used for features selection, to assess whether the activation patch for a specific residue already exhibits the target property (e.g., whether it is classified as a helix when steering towards helices); we proceed with intervention only if the target property is absent.

    Subsequently, we analyze the distribution of helices, strands, and coils, as well as the SASA within the generated protein backbones. When steering toward specific secondary structures, increasing $\lambda$ to positive values monotonically elevates the proportion of the target structure while reducing the others (Figure~\ref{fig:intervention_combined}, left). We extend this framework to physical properties by steering towards features associated with high solvent exposure. As illustrated in Figure~\ref{fig:intervention_combined} (right), we observe a direct correlation between $\lambda$ and the mean SASA of the generated backbones, demonstrating that FoldSAE can modulate not just local geometry but other protein properties. Notably, steering in order to increase exposure is more effective than toward decreasing it. This is consistent with the model's training distribution of natively folded, compact structures that already exhibit relatively low solvent accessibility.

\begin{figure}[t]
         \centering
         \begin{subfigure}[b]{0.48\textwidth}
             \centering
             \includegraphics[width=\linewidth]{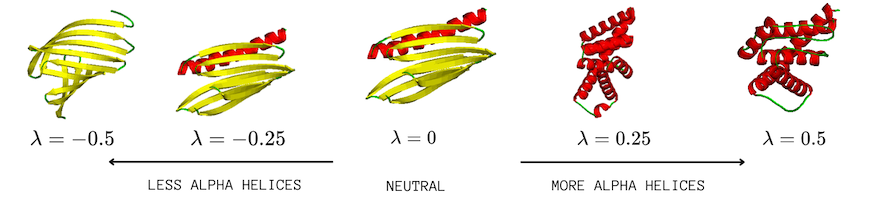}
             \caption{Steering secondary structure ($\lambda \in [-0.5, 0.5]$).}
             \label{fig:intervention_3d}
         \end{subfigure}
         \hfill
         \rule[-10pt]{0.1pt}{70pt} 
         \hfill
         \begin{subfigure}[b]{0.48\textwidth}
             \centering
             \includegraphics[width=\linewidth]{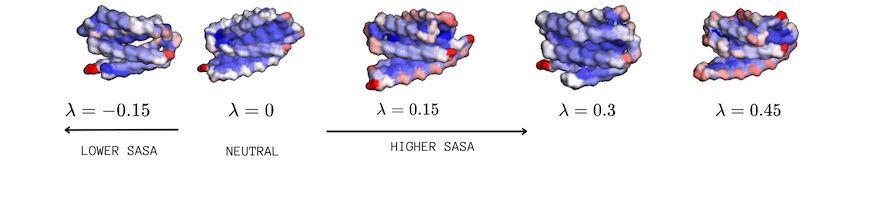}
             \caption{Steering SASA ($\lambda \in [-0.15, 0.45]$).}
             \label{fig:intervention_sasa}
         \end{subfigure}
    
         \caption{\textbf{Qualitative Analysis of Targeted Steering.} The figure displays generated backbones under varying steering intensities. Steering towards helices results in progressively more helical content (red ribbons). \textbf{(b)} Steering towards high SASA results in less compact, more exposed structures (visualized by surface potential).}
         \label{fig:intervention_single}
         \vspace{-1.5em}
    \end{figure}

    Moreover, we observe fine-grained control over the intervention even at the level of individual protein backbones. As shown in Figure~\ref{fig:intervention_single}, increasing $\lambda$ visibly alters the structural composition, resulting in a higher density of helices or a more expanded, accessible surface area depending on the steering target. More designs can be found in Appendix, Figure~\ref{fig:combined_visualisations}.

    \vspace{-1em}
    \paragraph{Validation of generated structures}
    To assess the biological plausibility of generated protein backbones following intervention, we compare their distribution against backbones generated without intervention. We embed backbones using ESM3~\citep{hayes2024simulating} and quantify distributional alignment using FBD~\citep{moller-larsen2025seqme} (an adaptation of FID~\citep{heusel2017gans} for proteins). To fairly evaluate structural quality independent of topological shifts, we weight reference proteins so that their distribution of helix-to-strand ratios (for secondary structure interventions) or SASA values (for SASA interventions) matches that of the corresponding generated batch. This ensures that any observed quality differences stem from the steering mechanism itself rather than from shifts in the overall structural composition.
    
    We independently evaluate batches of protein backbones across varying intervention strengths $\lambda$; the results are detailed in Appendix, Figure~\ref{fig:fbd_combined}. We observe that FBD scores for secondary structure interventions remain close to the neutral baseline ($\lambda=0$) across most of the steering range. This indicates that our mechanism preserves structural integrity and produces backbones resembling natural reference distributions. At higher intensities ($|\lambda| > 0.4$), the design success rate drops and FBD increases, aligning with the structural collapse observed in qualitative samples. For SASA interventions, the effective range of $\lambda$ where backbones remain intact is narrower, as extreme surface constraints may push the model toward biologically infeasible configurations. 
    
         

\begin{figure}[hbt]
 \centering
 \begin{subfigure}[b]{0.48\textwidth}
     \centering
     \includegraphics[width=\linewidth]{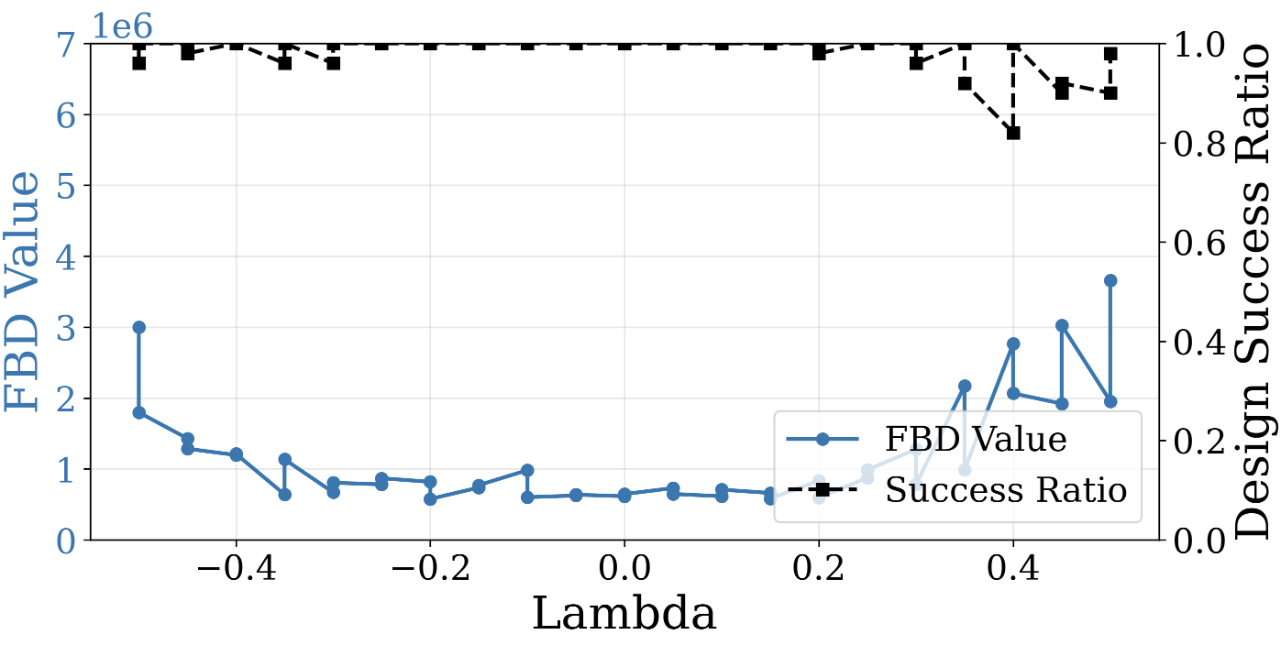}
     \caption{Secondary structure interventions.}
     \label{fig:fbd_struct}
 \end{subfigure}
 \begin{subfigure}[b]{0.48\textwidth}
     \centering
     \includegraphics[width=\linewidth]{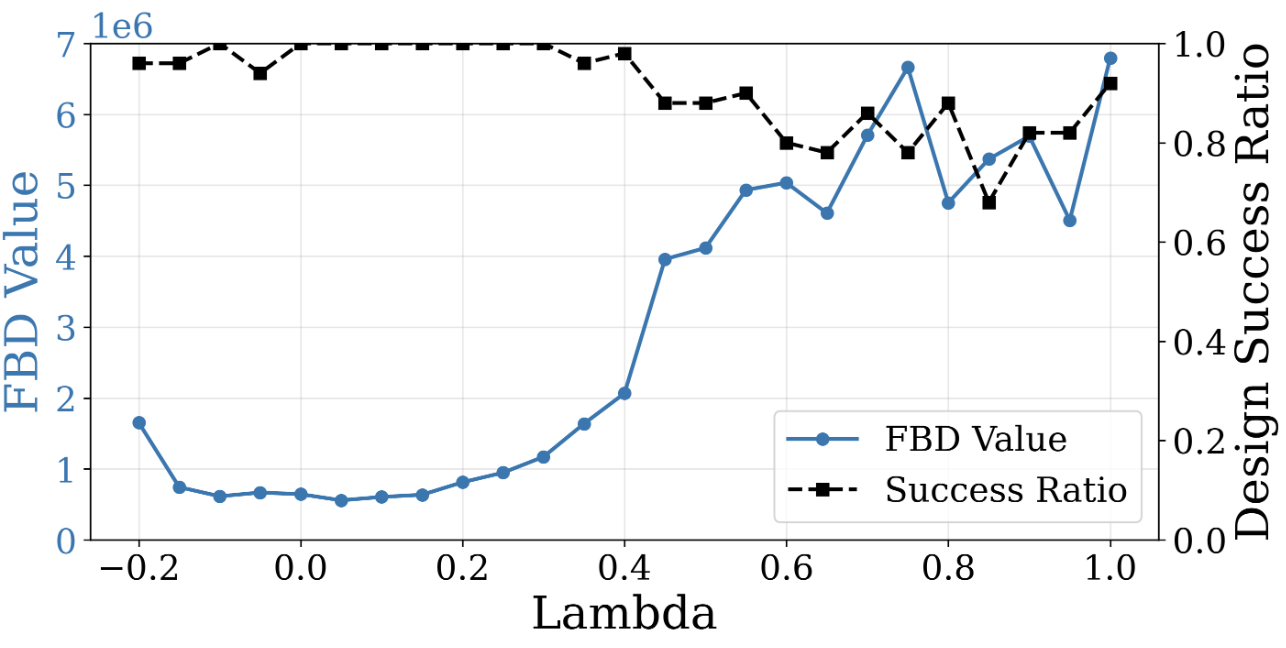}
     \caption{SASA interventions.}
     \label{fig:fbd_sasa}
 \end{subfigure}

     \caption{\textbf{Biological plausibility of steered backbones.} The Fréchet Biological Distance (FBD) scores (solid lines) and design success rates (dashed lines) are shown across a range of intervention strengths ($\lambda$) for both (a) secondary structure and (b) SASA interventions. FBD scores remain stable and close to the neutral baseline ($\lambda=0$) for moderate steering intensities, confirming that FoldSAE generates biologically realistic structures before extreme constraints induce structural collapse.}
     \label{fig:fbd_combined}
\end{figure}

\vspace{-1em}


\section{Conclusions}
\vspace{-1em}

In this work, we introduce \ours{}, a framework leveraging Sparse Autoencoders to decompose RFdiffusion's internal representations in a fully unsupervised manner. To demonstrate that discovered features are biologically meaningful and interpretable, we have shown that we can use SAE features selected from a single model to steer the generative process towards desired secondary structure formation or solvent accessible surface area (SASA). We note that the current analysis operates on protein backbones, and SASA estimates would benefit considerably from full atomic detail including side chains; however, the proposed framework is model-agnostic and readily extends to all-atom generative models as they become available.

\subsubsection*{Meaningfulness Statement}

By decomposing learned representations into sparse, interpretable features and enabling precise structural control, FoldSAE contributes to fundamental understanding of generative models for biology, with implications for reliable, interpretable design pipelines in clinical protein engineering.

\subsubsection*{Acknowledgments}

\noindent
\begin{minipage}{0.72\textwidth}
    This work was funded by the National Science Centre, Poland, grant no UMO-2023/51/B/ST6/03004 and the European Research Council (ERC) under the European Funding Union’s Horizon 2020 research and innovation programme (grant agreement No 810115 – DOG-AMP). The computing resources were provided by the PL-Grid Infrastructure, grant no.: PLG/2025/18390 and PLG/2025/18391.
\end{minipage}
\hfill
\begin{minipage}{0.25\textwidth}
    \centering
    \includegraphics[width=\linewidth, valign=m]{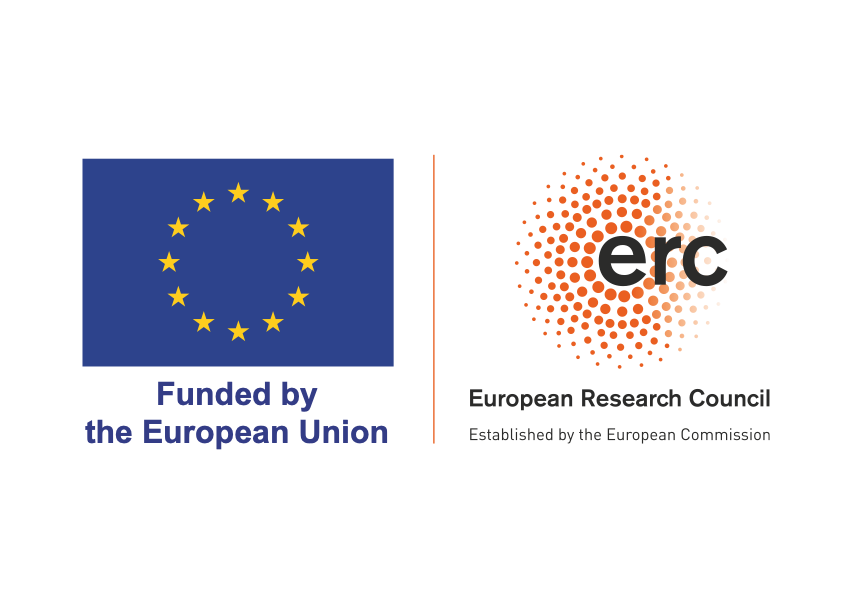}
\end{minipage}




\bibliography{iclr2025_conference}
\bibliographystyle{iclr2025_conference}

\appendix
    \section{Appendix}
\section{Results of intervention}
\begin{figure}
    \centering
    
    \begin{subfigure}{0.7\linewidth}
        \centering
        \includegraphics[width=\linewidth]{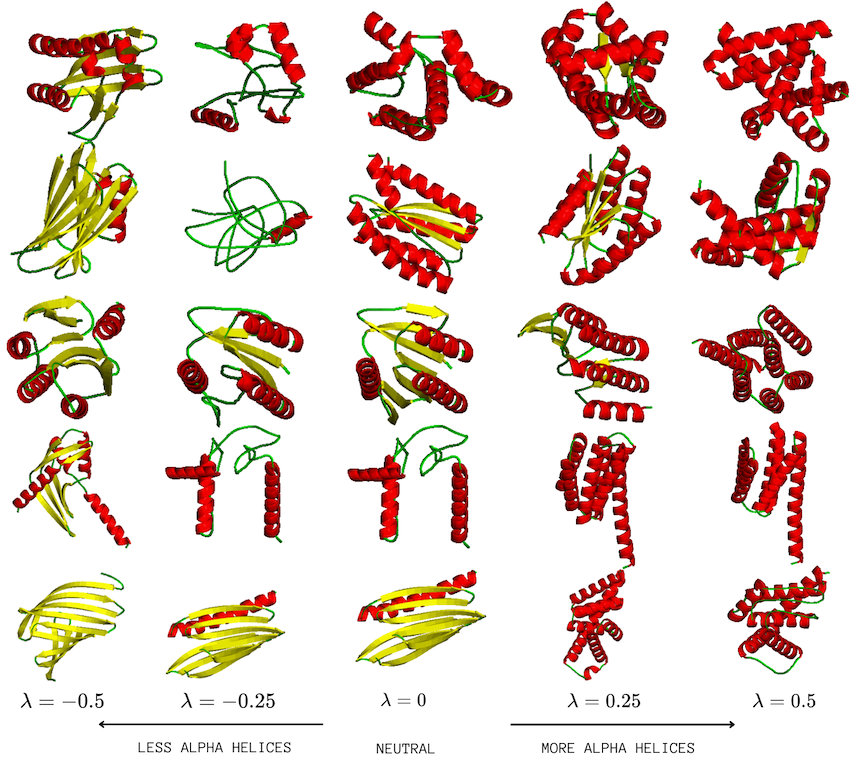}
        \caption{Backbone generation under varying steering intensities. Increased intensity correlates with higher helical content.}
        \label{fig:steering_visual}
    \end{subfigure}
    

    \begin{subfigure}{0.7\linewidth}
        \centering
        \includegraphics[width=\linewidth]{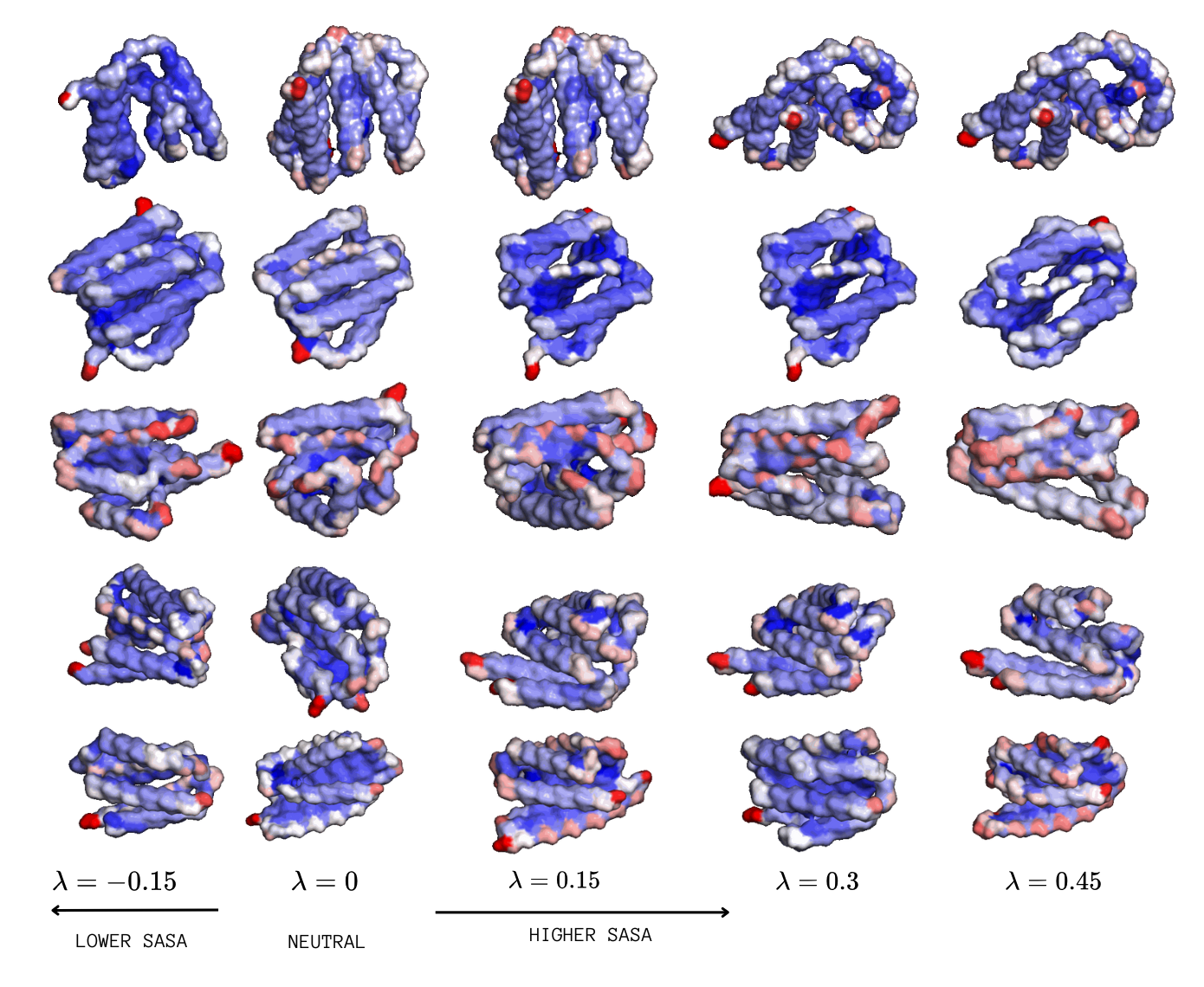}
        \caption{Backbone generation under varying steering intensities. Increased intensity correlates with brighter content.}
        \label{fig:sasa_visual}
    \end{subfigure}

    \caption{Visual analysis of generated structures: (a) illustrates the effect of steering on secondary structure distribution, while (b) shows the resulting surface properties/SASA.}
    \label{fig:combined_visualisations}
\end{figure}

\begin{figure}
    \centering
    \includegraphics[width=1\linewidth]{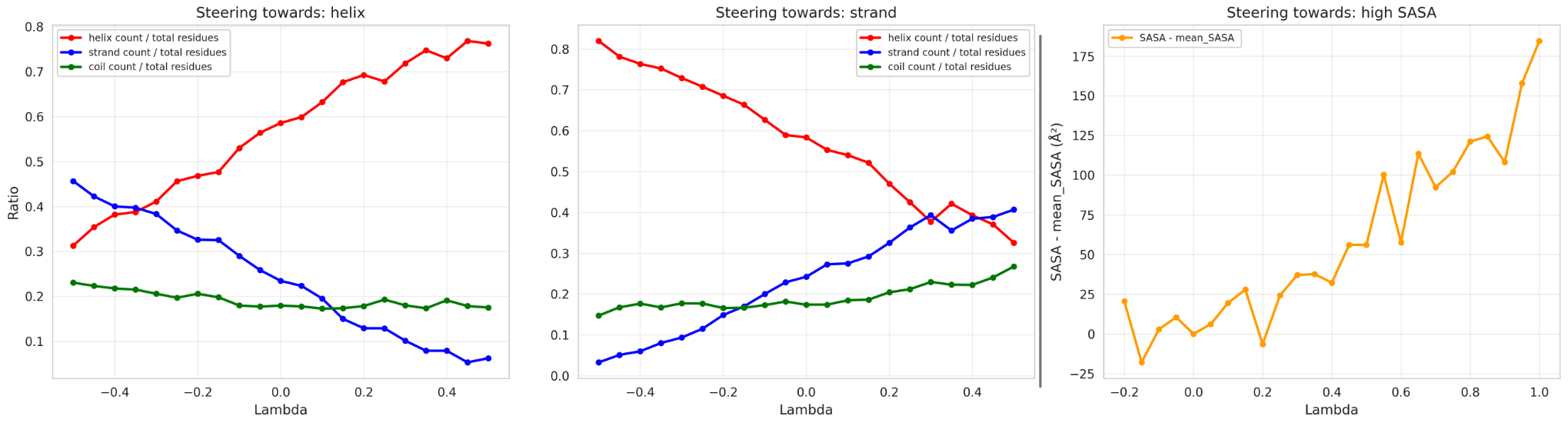}
    \caption{\textbf{FoldSAE intervention results.} (Left and Middle) Fraction of residues assigned to helices (red), strands (blue), and coils (green) as a function of steering intensity $\lambda$ when steering towards helices and strands, respectively. (Right) Solvent Accessible Surface Area (SASA) of generated backbones as a function of $\lambda$ when steering towards high-SASA features, showing a clear ability to modulate surface exposure.}
    \label{fig:scores_3_targets}
\end{figure}

Figure~\ref{fig:combined_visualisations} shows visualisation of targeted steering for multiple backbones. Figure~\ref{fig:scores_3_targets} shows quantitive results for steering towards helix, strand and high SASA side by side.

\section{Evaluated properties}
\label{app:labels}
While the proposed methodology is general and allows for unsupervised discovery of interpretable features, in this section, we propose to validate whether Sparse Autoencoder learns features that allow for differentiation between the final secondary structure of the generated backbone. Additionally, we evaluate solvent-accessible surface area (SASA) as a complementary, physically grounded measure of residue exposure that can be computed directly from generated structures.

\paragraph{Secondary structure} To annotate the generated backbones, we use STRIDE~\cite{frishman1995knowledge}, which assigns secondary structure classes to each residue based on hydrogen-bond energetics and torsion-angle propensities. We reduce the eight-state assignments to three states by mapping helical conformations (H, G, I) to helix, extended conformations (E, B) to strand, and all remaining states to coil, following the standard reduction scheme~\cite{rost1993prediction}. For evaluation, we measure ratio of given class to all residues.

\begin{figure}[!h]
    \centering
    \includegraphics[width=0.5\linewidth]{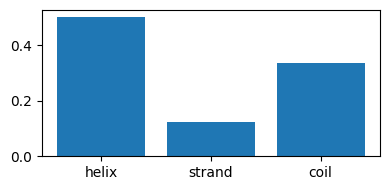}
    \caption{Distribution of helices, strands and coils in training dataset. We observe predominance of helices what matches distribtuion of secondary structures in natural proteins.}
    \label{fig:class_dist}
\end{figure}

 \label{app:classes_dist_uncond} We examine how frequent are helix, strand and coil in protein backbones generated in unconditional manner, as shown in Figure~\ref{fig:class_dist}.

\paragraph{SASA}We quantify residue exposure using solvent-accessible surface area (SASA), which measures the surface area of a biomolecule accessible to a solvent probe. We compute SASA using the Shrake-Rupley~\cite{} algorithm with FreeSASA~\cite{mitternacht2016freesasa}, an open-source implementation of standard SASA calculation routines. Since our method operates on generated backbones, we report backbone SASA computed from the atoms present in the structure and aggregate it across residues. When analyzing the relationship between SAE features and SASA, we operate on per-residue SASA.

\begin{figure}
    \centering
    \includegraphics[width=0.5\linewidth]{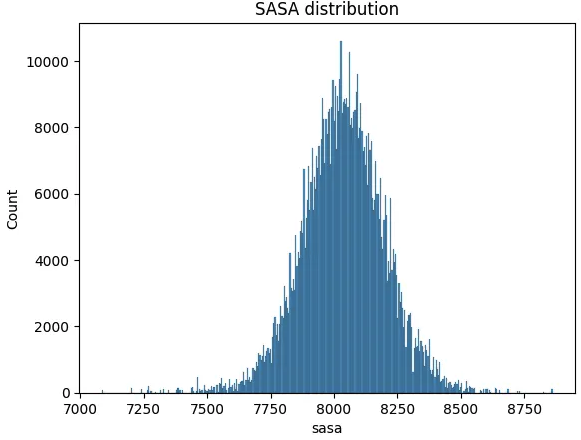}
    \caption{Distribution of SASA of proteins in training dataset. We observe normal distribution what support realism of generated backbones.}
    \label{fig:sasa_dist}
\end{figure}

We examine distribution of SASA in training datasets in Figure~\ref{fig:sasa_dist}.

\section{SAE in RFdiffusion}
    \paragraph{SAE training}
    \label{sae_training_experiments}

    We conducted a hyperparameter grid search for SAE training, sweeping across learning rates, expansion factors, and the sparsity parameter $k$; full details are provided in Appendix (see Table~\ref{tab:sae_grid}).
    Our final SAE was trained for 50,000 steps with a batch size of 4,096, employing a learning rate of $1 \times 10^{-4}$, an expansion factor of 16, and $k=64$.
    This configuration achieves an explained variance of 99.1\%, while maintaining a low fraction of both dead features (defined as latent neurons activating on fewer than 1 in $10^6$ training samples) and high-frequency features (activating on more than 1 in 100 examples).
    Minimizing the latter is crucial, as frequent features are prone to encoding multiple properties rather than being mono-semantic.
    The visualization of feature density is shown in Appendix, Figure~\ref{fig:sae_feature_density}.

    We gather a dataset for SAE training by collecting activations from chosen block for a set of 1200 protein backbones generated without any conditioning, for each timestep of diffusion process.
    To operate on single residue level, we flatten each collected activations vector and split it into $l$ patches, where $l$ denotes the number of residues in the protein. Then as described in Section~\ref{SAE_training_method} we train SAE to reconstruct activation patch for single residue. Table~\ref{tab:sae_grid} presents the full scope of our grid search across learning rates, expansion factors, and $k$ values from grid search described in Section~\ref{sae_training_experiments}.  Figure~\ref{fig:sae_feature_density} illustrates the distribution of feature activations in the final model, confirming a desirable sparsity profile with minimal dead or poly-semantic features. 
    

    \begin{figure}[h!]
        \centering
        \includegraphics[width=0.5\linewidth]{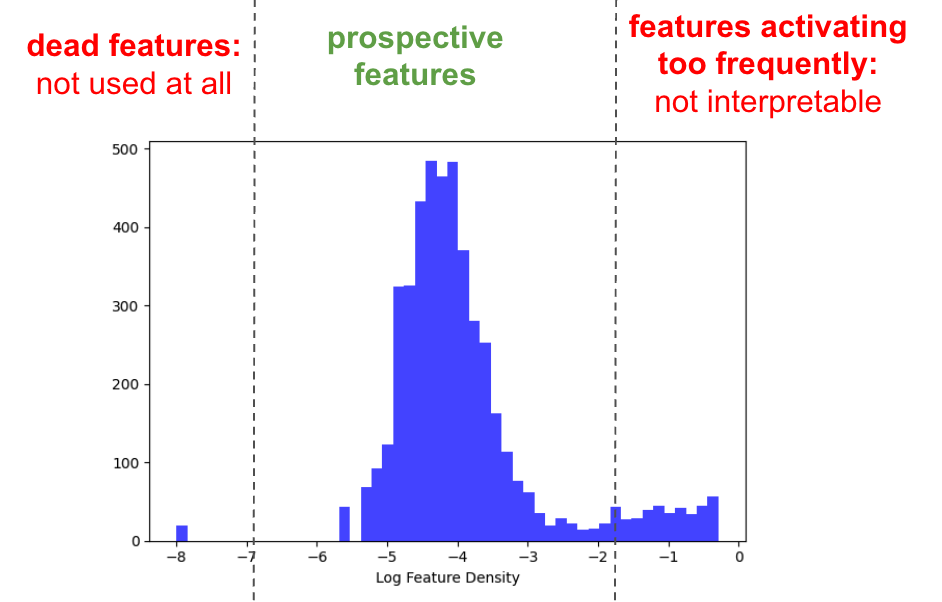}
        \caption{\textbf{Distribution of log feature density for the trained SAE.} The histogram illustrates the frequency of feature activations. The training setup results in a desirable distribution with a minimal fraction of dead features (left tail, $< 10^{-6}$) and high-frequency poly-semantic features (right tail, $> 10^{-2}$).}
        \label{fig:sae_feature_density}
    \end{figure}

    \begin{table}[!h]
\centering
\caption{\textbf{SAE training grid search.} We train SAE for various expansion factors (how many times latent space is wider), k in TopK, learning rate and report explained variance (the higher the better), ratio of dense features (sparsity below $10^{-2}$) (the lower the better) and ration of alive neuons (with sparsity below $10^{-5}$) (the higher the better).}
\label{tab:sae_grid}
\resizebox{\textwidth}{!}{
\begin{tabular}{cccccc|cccccc|cccccc|cccccc}
\toprule
Expl. Var. & Spars. $<10^{-5}$ & Spars. $<10^{-2}$ & Exp. & LR & k & Expl. Var. & Spars. $<10^{-5}$ & Spars. $<10^{-2}$ & Exp. & LR & k & Expl. Var. & Spars. $<10^{-5}$ & Spars. $<10^{-2}$ & Exp. & LR & k & Expl. Var. & Spars. $<10^{-5}$ & Spars. $<10^{-2}$ & Exp. & LR & k \\
\midrule
0.81 & 0.00e+00 & 0.47 & 32 & 0.0000 & 512 & 1.00 & 7.32e-01 & 0.88 & 8 & 0.0050 & 128 & 1.00 & 1.68e-01 & 0.84 & 32 & 0.0005 & 512 & 1.00 & 6.61e-01 & 0.87 & 8 & 0.0010 & 128 \\
1.00 & 8.46e-02 & 0.81 & 32 & 0.0005 & 512 & 1.00 & 7.37e-01 & 0.84 & 8 & 0.0010 & 128 & 0.99 & 0.00e+00 & 0.66 & 32 & 0.0001 & 512 & 0.75 & 8.53e-02 & 0.72 & 8 & 0.0000 & 64 \\
1.00 & 0.00e+00 & 0.67 & 32 & 0.0001 & 512 & 0.89 & 8.36e-02 & 0.79 & 8 & 0.0000 & 64 & 1.00 & 0.00e+00 & 0.00 & 32 & 0.0050 & 512 & 1.00 & 5.65e-01 & 0.93 & 8 & 0.0005 & 64 \\
1.00 & 7.17e-01 & 0.89 & 32 & 0.0050 & 512 & 1.00 & 6.93e-01 & 0.89 & 8 & 0.0005 & 64 & 1.00 & 2.85e-01 & 0.86 & 32 & 0.0010 & 512 & 0.99 & 8.02e-03 & 0.84 & 8 & 0.0001 & 64 \\
1.00 & 2.83e-01 & 0.84 & 32 & 0.0010 & 512 & 0.99 & 1.44e-02 & 0.84 & 8 & 0.0001 & 64 & 0.68 & 2.53e-03 & 0.64 & 32 & 0.0000 & 256 & 1.00 & 8.85e-01 & 0.95 & 8 & 0.0050 & 64 \\
0.88 & 6.33e-04 & 0.72 & 32 & 0.0000 & 256 & 1.00 & 9.03e-01 & 0.94 & 8 & 0.0050 & 64 & 1.00 & 3.79e-01 & 0.92 & 32 & 0.0005 & 256 & 1.00 & 7.47e-01 & 0.94 & 8 & 0.0010 & 64 \\
1.00 & 4.14e-01 & 0.89 & 32 & 0.0005 & 256 & 1.00 & 8.53e-01 & 0.92 & 8 & 0.0010 & 64 & 0.99 & 0.00e+00 & 0.81 & 32 & 0.0001 & 256 & 0.62 & 2.77e-01 & 0.90 & 8 & 0.0000 & 32 \\
1.00 & 0.00e+00 & 0.84 & 32 & 0.0001 & 256 & 0.80 & 4.28e-01 & 0.87 & 8 & 0.0000 & 32 & 1.00 & 0.00e+00 & 0.94 & 32 & 0.0050 & 256 & 1.00 & 6.25e-01 & 0.97 & 8 & 0.0005 & 32 \\
0.99 & 7.23e-01 & 0.93 & 32 & 0.0050 & 256 & 1.00 & 8.22e-01 & 0.94 & 8 & 0.0005 & 32 & 1.00 & 5.96e-01 & 0.93 & 32 & 0.0010 & 256 & 0.98 & 1.86e-01 & 0.94 & 8 & 0.0001 & 32 \\
1.00 & 6.84e-01 & 0.91 & 32 & 0.0010 & 256 & 0.98 & 2.63e-01 & 0.92 & 8 & 0.0001 & 32 & 0.66 & 4.11e-02 & 0.79 & 32 & 0.0000 & 128 & 1.00 & 8.79e-01 & 0.97 & 8 & 0.0050 & 32 \\
0.90 & 1.02e-02 & 0.85 & 32 & 0.0000 & 128 & 1.00 & 9.40e-01 & 0.96 & 8 & 0.0050 & 32 & 1.00 & 5.82e-01 & 0.96 & 32 & 0.0005 & 128 & 1.00 & 8.07e-01 & 0.97 & 8 & 0.0010 & 32 \\
1.00 & 7.38e-01 & 0.94 & 32 & 0.0005 & 128 & 1.00 & 8.96e-01 & 0.95 & 8 & 0.0010 & 32 & 0.99 & 1.06e-04 & 0.88 & 32 & 0.0001 & 128 & 0.98 & 0.00e+00 & 0.00 & 4 & 0.0000 & 512 \\
0.99 & 1.58e-03 & 0.92 & 32 & 0.0001 & 128 & 0.98 & 0.00e+00 & 0.00 & 4 & 0.0000 & 512 & 0.99 & 0.00e+00 & 0.97 & 32 & 0.0050 & 128 & 1.00 & 0.00e+00 & 0.05 & 4 & 0.0005 & 512 \\
1.00 & 9.55e-01 & 0.97 & 32 & 0.0050 & 128 & 1.00 & 0.00e+00 & 0.02 & 4 & 0.0005 & 512 & 1.00 & 8.05e-01 & 0.97 & 32 & 0.0010 & 128 & 1.00 & 0.00e+00 & 0.00 & 4 & 0.0001 & 512 \\
1.00 & 9.23e-01 & 0.96 & 32 & 0.0010 & 128 & 1.00 & 0.00e+00 & 0.00 & 4 & 0.0001 & 512 & 0.63 & 2.12e-01 & 0.90 & 32 & 0.0000 & 64 & 1.00 & 0.00e+00 & 0.00 & 4 & 0.0050 & 512 \\
0.90 & 1.60e-01 & 0.94 & 32 & 0.0000 & 64 & 1.00 & 0.00e+00 & 0.06 & 4 & 0.0050 & 512 & 1.00 & 7.49e-01 & 0.98 & 32 & 0.0005 & 64 & 1.00 & 8.45e-04 & 0.06 & 4 & 0.0010 & 512 \\
1.00 & 8.67e-01 & 0.97 & 32 & 0.0005 & 64 & 1.00 & 0.00e+00 & 0.06 & 4 & 0.0010 & 512 & 0.99 & 1.63e-02 & 0.95 & 32 & 0.0001 & 64 & 0.93 & 0.00e+00 & 0.02 & 4 & 0.0000 & 256 \\
0.99 & 7.07e-02 & 0.96 & 32 & 0.0001 & 64 & 0.96 & 0.00e+00 & 0.02 & 4 & 0.0000 & 256 & 1.00 & 9.38e-01 & 0.99 & 32 & 0.0050 & 64 & 1.00 & 2.79e-02 & 0.49 & 4 & 0.0005 & 256 \\
1.00 & 9.73e-01 & 0.98 & 32 & 0.0050 & 64 & 1.00 & 3.38e-03 & 0.46 & 4 & 0.0005 & 256 & 1.00 & 9.12e-01 & 0.98 & 32 & 0.0010 & 64 & 0.99 & 0.00e+00 & 0.15 & 4 & 0.0001 & 256 \\
1.00 & 9.44e-01 & 0.98 & 32 & 0.0010 & 64 & 0.99 & 0.00e+00 & 0.15 & 4 & 0.0001 & 256 & 0.58 & 4.11e-01 & 0.97 & 32 & 0.0000 & 32 & 1.00 & 4.39e-02 & 0.54 & 4 & 0.0050 & 256 \\
0.82 & 6.83e-01 & 0.97 & 32 & 0.0000 & 32 & 1.00 & 4.22e-02 & 0.53 & 4 & 0.0050 & 256 & 1.00 & 8.63e-01 & 0.99 & 32 & 0.0005 & 32 & 1.00 & 2.12e-01 & 0.52 & 4 & 0.0010 & 256 \\
1.00 & 9.30e-01 & 0.98 & 32 & 0.0005 & 32 & 1.00 & 2.20e-01 & 0.51 & 4 & 0.0010 & 256 & 0.98 & 2.52e-01 & 0.98 & 32 & 0.0001 & 32 & 0.87 & 0.00e+00 & 0.21 & 4 & 0.0000 & 128 \\
0.98 & 6.25e-01 & 0.98 & 32 & 0.0001 & 32 & 0.93 & 0.00e+00 & 0.29 & 4 & 0.0000 & 128 & 1.00 & 9.72e-01 & 0.99 & 32 & 0.0050 & 32 & 1.00 & 2.10e-01 & 0.76 & 4 & 0.0005 & 128 \\
1.00 & 9.82e-01 & 0.99 & 32 & 0.0050 & 32 & 1.00 & 1.41e-01 & 0.71 & 4 & 0.0005 & 128 & 1.00 & 9.46e-01 & 0.99 & 32 & 0.0010 & 32 & 0.99 & 0.00e+00 & 0.41 & 4 & 0.0001 & 128 \\
1.00 & 9.67e-01 & 0.99 & 32 & 0.0010 & 32 & 0.99 & 0.00e+00 & 0.52 & 4 & 0.0001 & 128 & 0.83 & 0.00e+00 & 0.19 & 16 & 0.0000 & 512 & 1.00 & 2.71e-01 & 0.76 & 4 & 0.0050 & 128 \\
0.88 & 0.00e+00 & 0.19 & 16 & 0.0000 & 512 & 1.00 & 6.17e-01 & 0.77 & 4 & 0.0050 & 128 & 1.00 & 7.09e-02 & 0.71 & 16 & 0.0005 & 512 & 1.00 & 4.25e-01 & 0.76 & 4 & 0.0010 & 128 \\
1.00 & 2.13e-02 & 0.68 & 16 & 0.0005 & 512 & 1.00 & 4.18e-01 & 0.74 & 4 & 0.0010 & 128 & 0.99 & 0.00e+00 & 0.46 & 16 & 0.0001 & 512 & 0.76 & 4.31e-02 & 0.53 & 4 & 0.0000 & 64 \\
0.99 & 0.00e+00 & 0.42 & 16 & 0.0001 & 512 & 0.88 & 3.97e-02 & 0.60 & 4 & 0.0000 & 64 & 1.00 & 0.00e+00 & 0.00 & 16 & 0.0050 & 512 & 1.00 & 4.44e-01 & 0.87 & 4 & 0.0005 & 64 \\
1.00 & 1.72e-01 & 0.77 & 16 & 0.0050 & 512 & 1.00 & 5.44e-01 & 0.81 & 4 & 0.0005 & 64 & 1.00 & 1.76e-01 & 0.74 & 16 & 0.0010 & 512 & 0.99 & 8.45e-04 & 0.73 & 4 & 0.0001 & 64 \\
1.00 & 1.07e-01 & 0.71 & 16 & 0.0010 & 512 & 0.99 & 8.45e-04 & 0.73 & 4 & 0.0001 & 64 & 0.80 & 1.27e-03 & 0.44 & 16 & 0.0000 & 256 & 1.00 & 7.45e-01 & 0.89 & 4 & 0.0050 & 64 \\
0.92 & 0.00e+00 & 0.51 & 16 & 0.0000 & 256 & 1.00 & 7.86e-01 & 0.88 & 4 & 0.0050 & 64 & 1.00 & 2.92e-01 & 0.86 & 16 & 0.0005 & 256 & 1.00 & 5.98e-01 & 0.89 & 4 & 0.0010 & 64 \\
1.00 & 2.71e-01 & 0.81 & 16 & 0.0005 & 256 & 1.00 & 7.32e-01 & 0.84 & 4 & 0.0010 & 64 & 0.99 & 0.00e+00 & 0.68 & 16 & 0.0001 & 256 & 0.61 & 1.79e-01 & 0.80 & 4 & 0.0000 & 32 \\
1.00 & 0.00e+00 & 0.71 & 16 & 0.0001 & 256 & 0.79 & 2.68e-01 & 0.74 & 4 & 0.0000 & 32 & 1.00 & 1.82e-01 & 0.88 & 16 & 0.0050 & 256 & 1.00 & 4.84e-01 & 0.93 & 4 & 0.0005 & 32 \\
1.00 & 6.14e-01 & 0.88 & 16 & 0.0050 & 256 & 1.00 & 7.09e-01 & 0.89 & 4 & 0.0005 & 32 & 1.00 & 6.50e-01 & 0.88 & 16 & 0.0010 & 256 & 0.98 & 1.01e-01 & 0.90 & 4 & 0.0001 & 32 \\
1.00 & 6.21e-01 & 0.83 & 16 & 0.0010 & 256 & 0.98 & 1.11e-01 & 0.85 & 4 & 0.0001 & 32 & 0.75 & 2.77e-02 & 0.66 & 16 & 0.0000 & 128 & 1.00 & 7.09e-01 & 0.94 & 4 & 0.0050 & 32 \\
0.92 & 7.81e-03 & 0.72 & 16 & 0.0000 & 128 & 1.00 & 8.90e-01 & 0.93 & 4 & 0.0050 & 32 & 1.00 & 5.29e-01 & 0.93 & 16 & 0.0005 & 128 & 1.00 & 6.22e-01 & 0.94 & 4 & 0.0010 & 32 \\
1.00 & 6.91e-01 & 0.89 & 16 & 0.0005 & 128 & 1.00 & 8.36e-01 & 0.91 & 4 & 0.0010 & 32 & 0.99 & 0.00e+00 & 0.80 & 16 & 0.0001 & 128 & 0.96 & 0.00e+00 & 0.00 & 2 & 0.0000 & 512 \\
0.99 & 4.22e-04 & 0.85 & 16 & 0.0001 & 128 & 0.96 & 0.00e+00 & 0.00 & 2 & 0.0000 & 512 & 1.00 & 8.22e-01 & 0.95 & 16 & 0.0050 & 128 & 1.00 & 0.00e+00 & 0.00 & 2 & 0.0005 & 512 \\
1.00 & 8.86e-01 & 0.94 & 16 & 0.0050 & 128 & 1.00 & 0.00e+00 & 0.00 & 2 & 0.0005 & 512 & 1.00 & 8.06e-01 & 0.94 & 16 & 0.0010 & 128 & 0.99 & 0.00e+00 & 0.00 & 2 & 0.0001 & 512 \\
1.00 & 8.48e-01 & 0.91 & 16 & 0.0010 & 128 & 0.99 & 0.00e+00 & 0.00 & 2 & 0.0001 & 512 & 0.71 & 1.43e-01 & 0.82 & 16 & 0.0000 & 64 & 1.00 & 0.00e+00 & 0.00 & 2 & 0.0050 & 512 \\
0.89 & 1.26e-01 & 0.89 & 16 & 0.0000 & 64 & 1.00 & 0.00e+00 & 0.00 & 2 & 0.0050 & 512 & 1.00 & 6.52e-01 & 0.97 & 16 & 0.0005 & 64 & 1.00 & 0.00e+00 & 0.00 & 2 & 0.0010 & 512 \\
1.00 & 8.68e-01 & 0.95 & 16 & 0.0005 & 64 & 1.00 & 0.00e+00 & 0.00 & 2 & 0.0010 & 512 & 0.99 & 6.55e-03 & 0.89 & 16 & 0.0001 & 64 & 0.95 & 0.00e+00 & 0.00 & 2 & 0.0000 & 256 \\
0.99 & 7.35e-02 & 0.91 & 16 & 0.0001 & 64 & 0.96 & 0.00e+00 & 0.00 & 2 & 0.0000 & 256 & 1.00 & 9.16e-01 & 0.97 & 16 & 0.0050 & 64 & 1.00 & 0.00e+00 & 0.02 & 2 & 0.0005 & 256 \\
1.00 & 9.51e-01 & 0.97 & 16 & 0.0050 & 64 & 1.00 & 0.00e+00 & 0.02 & 2 & 0.0005 & 256 & 1.00 & 8.63e-01 & 0.97 & 16 & 0.0010 & 64 & 0.99 & 0.00e+00 & 0.00 & 2 & 0.0001 & 256 \\
1.00 & 9.21e-01 & 0.96 & 16 & 0.0010 & 64 & 0.99 & 0.00e+00 & 0.00 & 2 & 0.0001 & 256 & 0.61 & 3.89e-01 & 0.94 & 16 & 0.0000 & 32 & 1.00 & 0.00e+00 & 0.07 & 2 & 0.0050 & 256 \\
0.81 & 5.84e-01 & 0.94 & 16 & 0.0000 & 32 & 1.00 & 0.00e+00 & 0.05 & 2 & 0.0050 & 256 & 1.00 & 8.53e-01 & 0.98 & 16 & 0.0005 & 32 & 1.00 & 0.00e+00 & 0.05 & 2 & 0.0010 & 256 \\
1.00 & 8.78e-01 & 0.97 & 16 & 0.0005 & 32 & 1.00 & 0.00e+00 & 0.03 & 2 & 0.0010 & 256 & 0.98 & 2.69e-01 & 0.97 & 16 & 0.0001 & 32 & 0.87 & 1.69e-03 & 0.04 & 2 & 0.0000 & 128 \\
0.98 & 4.52e-01 & 0.96 & 16 & 0.0001 & 32 & 0.92 & 0.00e+00 & 0.05 & 2 & 0.0000 & 128 & 1.00 & 9.44e-01 & 0.98 & 16 & 0.0050 & 32 & 1.00 & 2.36e-02 & 0.48 & 2 & 0.0005 & 128 \\
1.00 & 9.71e-01 & 0.98 & 16 & 0.0050 & 32 & 1.00 & 4.39e-02 & 0.45 & 2 & 0.0005 & 128 & 1.00 & 8.95e-01 & 0.98 & 16 & 0.0010 & 32 & 0.99 & 0.00e+00 & 0.16 & 2 & 0.0001 & 128 \\
1.00 & 9.27e-01 & 0.97 & 16 & 0.0010 & 32 & 0.99 & 0.00e+00 & 0.16 & 2 & 0.0001 & 128 & 0.93 & 0.00e+00 & 0.02 & 8 & 0.0000 & 512 & 1.00 & 5.41e-02 & 0.53 & 2 & 0.0050 & 128 \\
0.95 & 0.00e+00 & 0.02 & 8 & 0.0000 & 512 & 1.00 & 8.61e-02 & 0.56 & 2 & 0.0050 & 128 & 1.00 & 2.11e-03 & 0.49 & 8 & 0.0005 & 512 & 1.00 & 1.45e-01 & 0.52 & 2 & 0.0010 & 128 \\
1.00 & 4.22e-04 & 0.44 & 8 & 0.0005 & 512 & 1.00 & 2.20e-01 & 0.49 & 2 & 0.0010 & 128 & 0.99 & 0.00e+00 & 0.20 & 8 & 0.0001 & 512 & 0.74 & 1.69e-02 & 0.23 & 2 & 0.0000 & 64 \\
1.00 & 0.00e+00 & 0.16 & 8 & 0.0001 & 512 & 0.86 & 1.18e-02 & 0.25 & 2 & 0.0000 & 64 & 1.00 & 0.00e+00 & 0.00 & 8 & 0.0050 & 512 & 1.00 & 1.82e-01 & 0.75 & 2 & 0.0005 & 64 \\
1.00 & 3.68e-01 & 0.57 & 8 & 0.0050 & 512 & 1.00 & 2.92e-01 & 0.71 & 2 & 0.0005 & 64 & 1.00 & 7.60e-02 & 0.51 & 8 & 0.0010 & 512 & 0.98 & 1.69e-03 & 0.51 & 2 & 0.0001 & 64 \\
1.00 & 1.31e-02 & 0.48 & 8 & 0.0010 & 512 & 0.99 & 0.00e+00 & 0.53 & 2 & 0.0001 & 64 & 0.87 & 0.00e+00 & 0.21 & 8 & 0.0000 & 256 & 1.00 & 3.97e-01 & 0.77 & 2 & 0.0050 & 64 \\
0.94 & 0.00e+00 & 0.25 & 8 & 0.0000 & 256 & 1.00 & 5.42e-01 & 0.76 & 2 & 0.0050 & 64 & 1.00 & 2.55e-01 & 0.75 & 8 & 0.0005 & 256 & 1.00 & 3.50e-01 & 0.76 & 2 & 0.0010 & 64 \\
1.00 & 8.99e-02 & 0.70 & 8 & 0.0005 & 256 & 1.00 & 5.81e-01 & 0.72 & 2 & 0.0010 & 64 & 0.99 & 0.00e+00 & 0.45 & 8 & 0.0001 & 256 & 0.56 & 8.28e-02 & 0.53 & 2 & 0.0000 & 32 \\
1.00 & 0.00e+00 & 0.50 & 8 & 0.0001 & 256 & 0.77 & 7.94e-02 & 0.49 & 2 & 0.0000 & 32 & 1.00 & 3.18e-01 & 0.77 & 8 & 0.0050 & 256 & 1.00 & 2.82e-01 & 0.86 & 2 & 0.0005 & 32 \\
0.99 & 3.67e-01 & 0.71 & 8 & 0.0050 & 256 & 1.00 & 4.70e-01 & 0.81 & 2 & 0.0005 & 32 & 1.00 & 5.25e-01 & 0.76 & 8 & 0.0010 & 256 & 0.97 & 3.21e-02 & 0.75 & 2 & 0.0001 & 32 \\
1.00 & 4.21e-01 & 0.73 & 8 & 0.0010 & 256 & 0.98 & 2.36e-02 & 0.73 & 2 & 0.0001 & 32 & 0.81 & 9.71e-03 & 0.47 & 8 & 0.0000 & 128 & 1.00 & 6.03e-01 & 0.87 & 2 & 0.0050 & 32 \\
0.93 & 2.53e-03 & 0.55 & 8 & 0.0000 & 128 & 1.00 & 7.84e-01 & 0.85 & 2 & 0.0050 & 32 & 1.00 & 5.19e-01 & 0.87 & 8 & 0.0005 & 128 & 1.00 & 5.15e-01 & 0.87 & 2 & 0.0010 & 32 \\
1.00 & 3.94e-01 & 0.83 & 8 & 0.0005 & 128 & 1.00 & 7.15e-01 & 0.82 & 2 & 0.0010 & 32 & 0.99 & 0.00e+00 & 0.64 & 8 & 0.0001 & 128 &  &  &  &  &  &  \\
1.00 & 0.00e+00 & 0.74 & 8 & 0.0001 & 128 & 0.66 & 0.00e+00 & 0.42 & 32 & 0.0000 & 512 & 1.00 & 6.68e-01 & 0.89 & 8 & 0.0050 & 128 &  &  &  &  &  &  \\
\bottomrule
\end{tabular}

}
\end{table}

    \paragraph{SAE at intervention}
    When intervention is required, we must address the fact that reconstructing activations with SAE inherently introduces reconstruction error. If propagated to subsequent blocks, this error causes a distribution shift in activations that can degrade downstream performance. To mitigate this, we offset the error using the following procedure:
    \begin{enumerate}
        \item We reconstruct the original activations $\Gamma$ without applying any intervention, yielding $\hat{\Gamma}$.
        \item We then calculate the original reconstruction error as $E = \hat{\Gamma} - \Gamma$.
        \item We reconstruct the activations $\Gamma$ \textit{while applying} the intervention, which results in $\hat{\Gamma}'$.
        \item The final activations returned after the intervention are the intervened reconstruction offset by the original error: $\hat{\Gamma}' + E$.
    \end{enumerate}

\section{Ablation of RFdiffusion blocks}

\begin{figure}[H]
        \centering
        \includegraphics[width=1.0\linewidth]{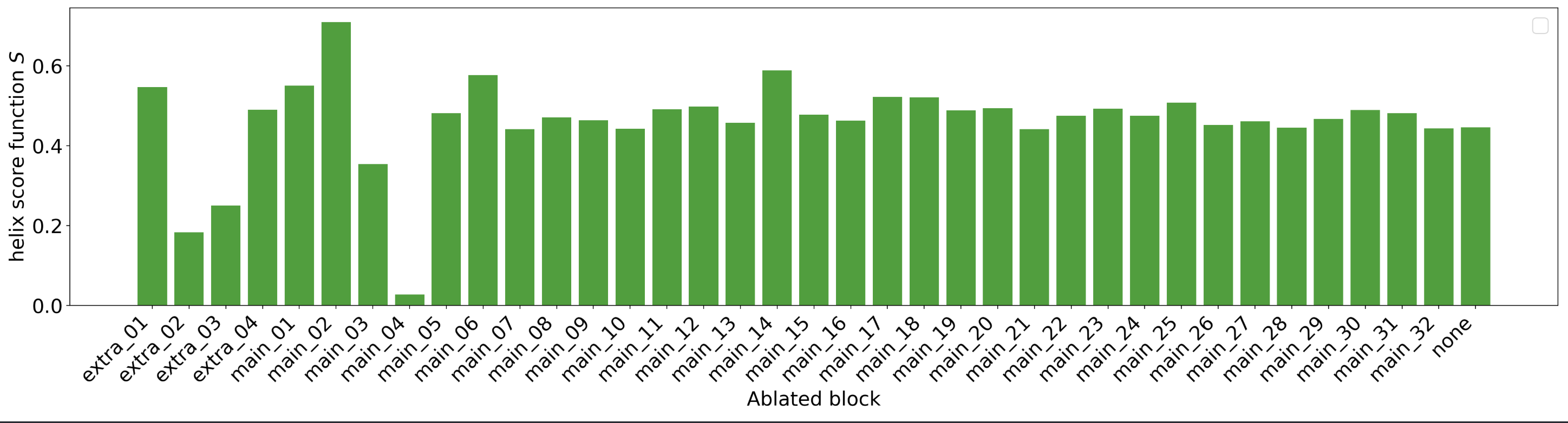}
        \caption{\textbf{Localization of secondary structure encoding.} Impact of systematic block ablation on secondary structure generation. The bar plot displays the helix score (fraction of residues in helices) for each configuration. Ablation of block $\textit{main\_04}$ causes a collapse in helix generation, identifying it as the critical component for this property.}
        \label{fig:ablation_results}
    \end{figure}

\begin{figure}[H]
        \centering
        \includegraphics[width=1.0\linewidth]{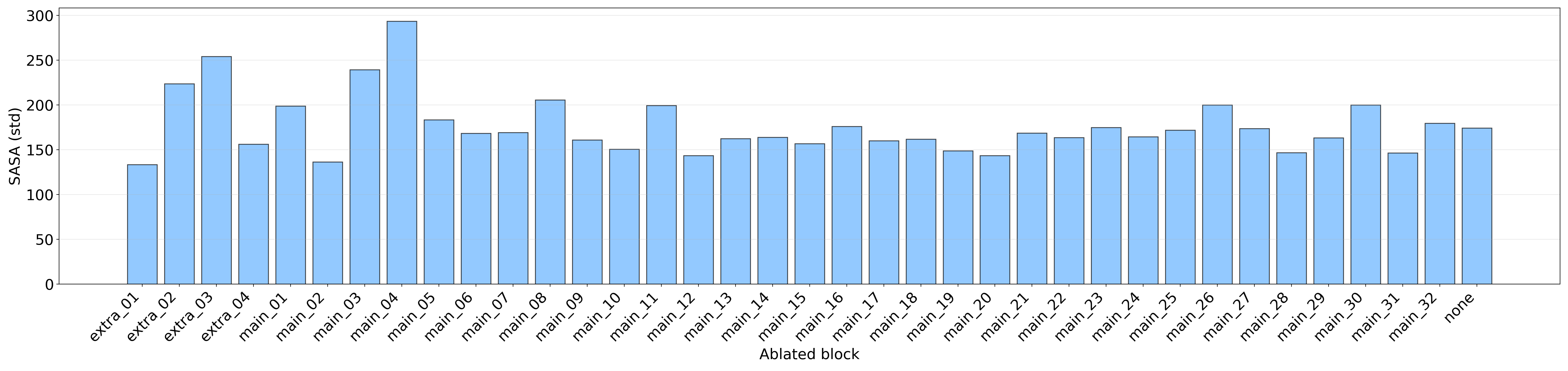}
        \caption{\textbf{Localization of SASA encoding.} Impact of systematic block ablation on Solvent Accessible Surface Area (SASA). The bar plot displays the SASA score function (standard deviation) for each configuration. Ablating block $\textit{main\_04}$ results in the highest variance, indicating a loss of control over surface properties.}
        \label{fig:ablation_results_sasa}
    \end{figure}

We systematically ablate blocks in the RFdiffusion to localize the semantic encoding of our target properties.
    We measure the degradation of the generative capability by comparing the helix score (Figure~\ref{fig:ablation_results}) and SASA distribution statistics (Figure~\ref{fig:ablation_results_sasa}) of ablated models against the original.
    In both cases, block $\textit{main\_04}$ is identified as the most critical component; its removal causes the most significant divergence from the baseline behavior, maximizing the score difference described at Section~\ref{blocks_ablation}.

\section{Feature selection} \label{probing_detials}
    \paragraph{Dataset creation}
    We gather a new dataset for the training of probing models and the analysis of their coefficients. First, using RFdiffusion with integrated SAE we generate 10000 of proteins without making any interventions and store the SAE encoder's activations together with their associated timesteps, proteins and residues. Using Stride, to each residue we assign secondary structure, and map these assignments to corresponding SAE activations. SASA annotations are on residue level using~\cite{mitternacht2016freesasa}. We visualize this approach in Appendix, (Figure~\ref{fig:dataset_gathering}). \paulina{sth wrong here}

    \begin{figure}[h!]
        \centering
        \includegraphics[width=1.0\linewidth]{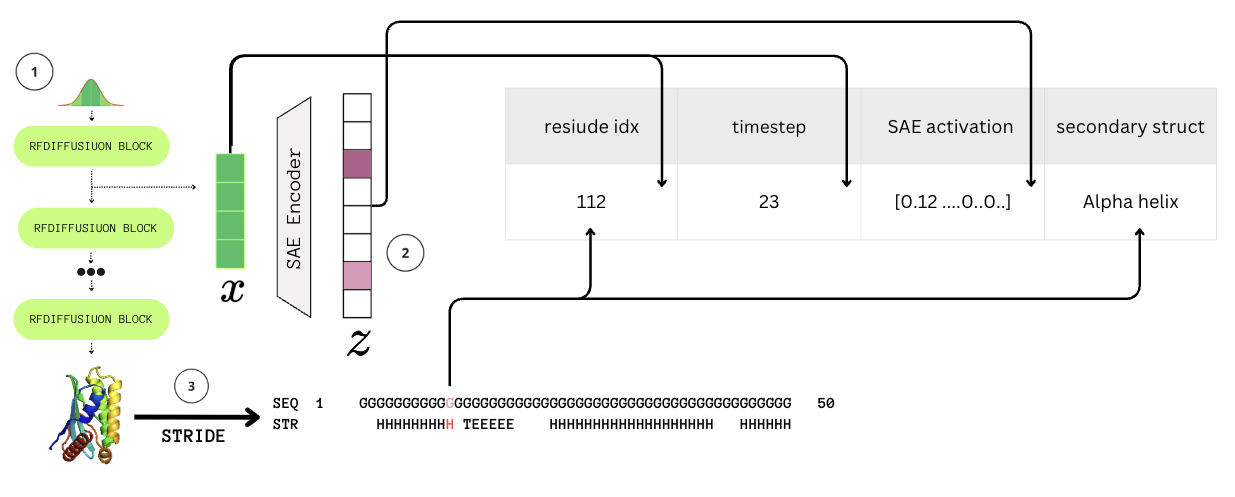}
        \caption{\textbf{Dataset creation}. 1) We generate proteins and cache activations of chosen block from each timestep, spliting them into patches per residue. 2) We reconstruct block activations with SAE and cache SAE encoder activations for each residue. 3) We assign secondary structure for each residue 
        }
        \label{fig:dataset_gathering}
    \end{figure}

    \paragraph{Probing model training} 
    We develop binary classifiers in a One-vs-Rest (OvR) framework.
    For secondary structures, we train separate `helix vs.\ rest' and `strand vs.\ rest' models. To mitigate the class imbalance present in secondary structures (Figure~\ref{fig:class_dist}), we apply class weighting during optimization.

    In case of SASA we address it's continuous nature by discretizing the target by training classifiers to distinguish extreme quartiles: `above $3^{\text{rd}}$ quartile ($>Q3$) vs.\ rest' and `below $1^{\text{st}}$ quartile ($<Q1$) vs.\ rest'.
    We evaluate these classifiers in two configurations: \textit{time-dependent} (trained on activations from specific timesteps) and \textit{time-agnostic} (trained on activations pooled across all steps).

\begin{figure}[H]
        \centering
        \includegraphics[width=1\linewidth]{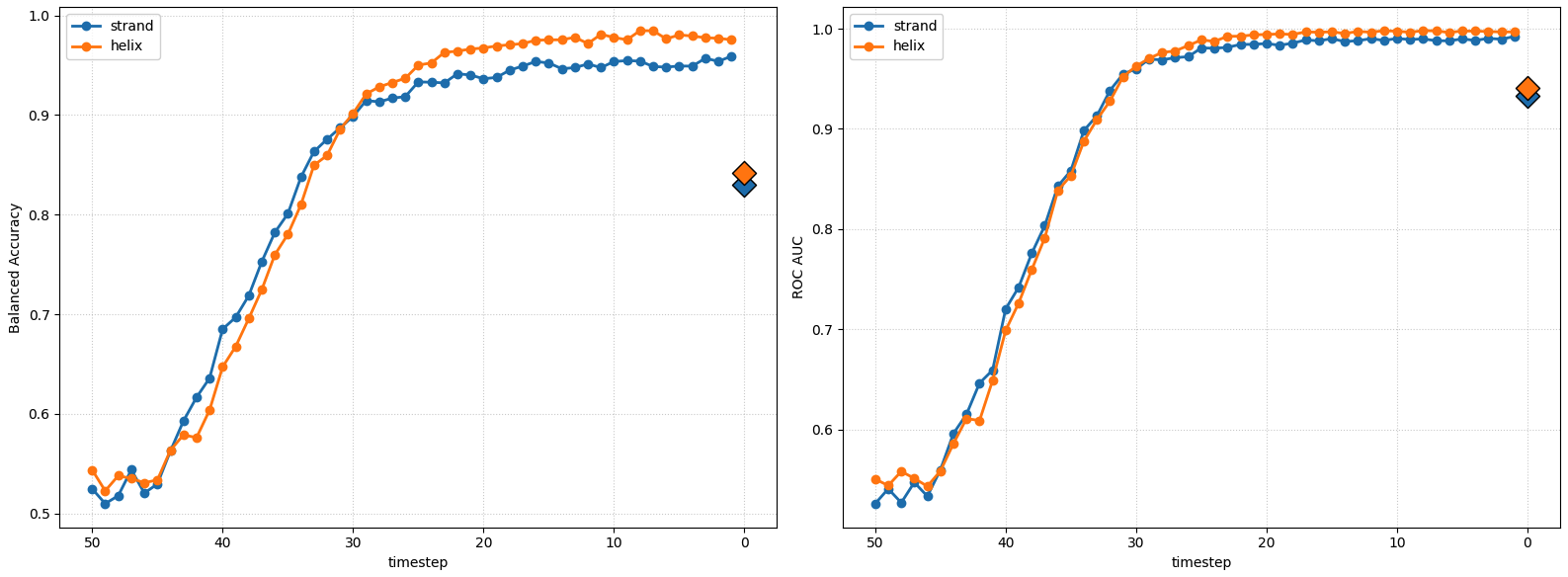}
        \caption{Results of training \textit{time-agnostic} and \textit{time-dependent} probing models. The first diffusion step is $50$ and the last is $1$; the diamond at $0$ denotes the score for the \textit{time-agnostic} model. The left pane reports balanced accuracy and the right pane reports AUC ROC. We observe robust performance of \textit{time-agnostic models} compared to individual timesteps.}
        \label{fig:apppendix_probes_timesteps}
    \end{figure}

    Notably, the models maintain robust performance even in the \textit{time-agnostic} setting.
    We achieve ROC AUC scores of $94.1\%$ and $93.3\%$ for the helix and strand tasks, respectively.
    The SASA probes also perform reliably, reaching $81.78\%$ AUC for the high-SASA ($>Q3$) detector and $78.86\%$ for the low-SASA ($<Q1$) detector.
    In this section, we provide the detailed breakdown of classifier performance across the diffusion trajectory, complementing the summary statistics provided in the main text. Figure~\ref{fig:apppendix_probes_timesteps} illustrates the stability of the probing models by comparing the \textit{time-dependent} performance at each diffusion step ($t=50 \to 1$) against the \textit{time-agnostic} baseline.
    
    \paragraph{Features selection based on probing models}

    \begin{figure}[h!]
        \centering
        \includegraphics[width=1\linewidth]{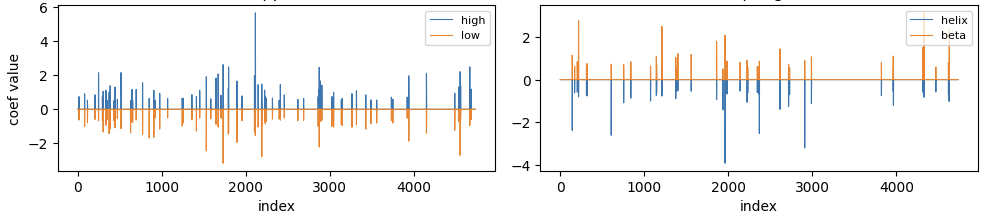}
        \caption{\textbf{FoldSAE interpretation.} Visualization of regression coefficients for the high (above 3rd quartile) classifiers - red and low (below 1str quartile) classifier - blue. Coefficients with an absolute magnitude greater than or equal to 0.1 are highlighted as solid lines, while those with magnitudes less than 0.1 are depicted as faint dashed lines. The largest coefficients often coincide at the same feature indices but exhibit opposite signs, suggesting a shared set of latent features governs the structural differentiation.}
        \label{fig:probes_all_coefs}
    \end{figure}

    \begin{figure}[h!]
        \centering
        \includegraphics[width=1\linewidth]{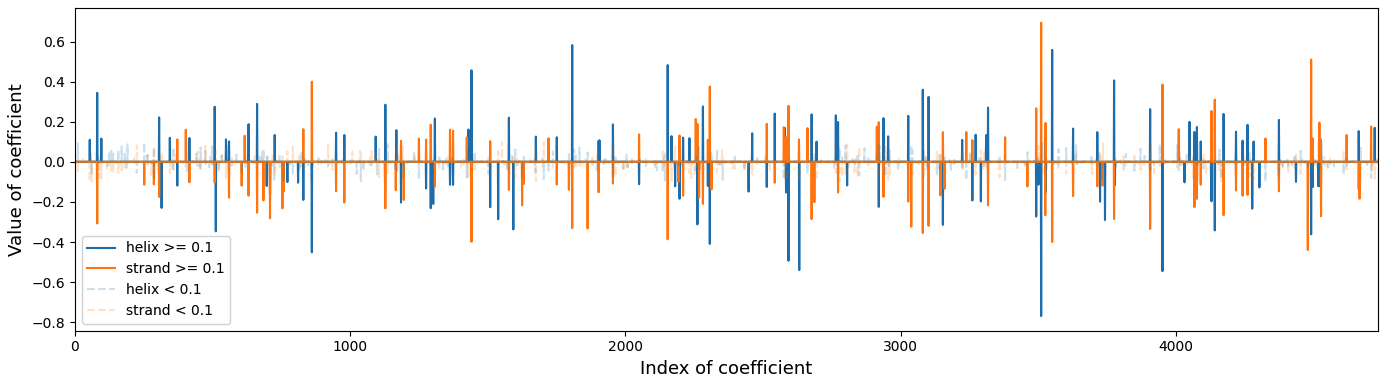}
        \caption{\textbf{FoldSAE interpretation.} Visualization of regression coefficients for the helix vs. rest (blue) and strand vs. rest (orange) probing classifiers. Coefficients with an absolute magnitude greater than or equal to 0.1 are highlighted as solid lines, while those with magnitudes less than 0.1 are depicted as faint dashed lines. The largest coefficients often coincide at the same feature indices but exhibit opposite signs, suggesting a shared set of latent features governs the structural differentiation.}
        \label{fig:probes_all_coefs}
    \end{figure}

\begin{figure}
        \centering
        \includegraphics[width=0.75\linewidth]{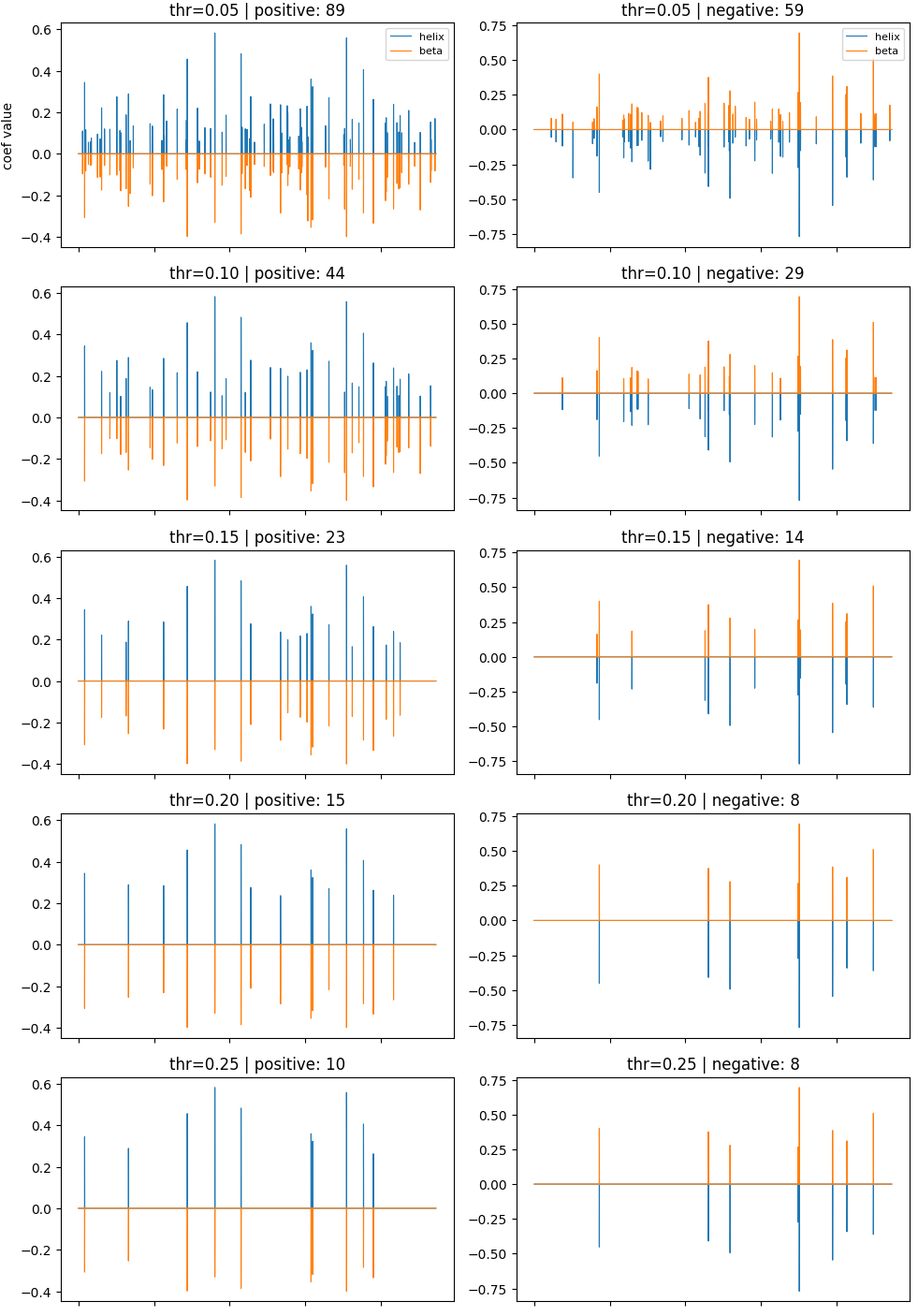}
        \caption{Visualization of the feature selection process using probing model coefficients. The panels illustrate the sparsification of features as the selection threshold increases (rows, from 0.05 to 0.35). Blue and orange bars represent coefficients for \textit{alpha helix} and \textit{beta sheet} classes, respectively. The subplot titles display the count of remaining discriminative features that satisfy two conditions: the coefficient modulus exceeds the chosen threshold, and the coefficients for the two classes exhibit opposite signs.}    \label{fig:appendix_probes_coefs_analysis}
    \end{figure}

    One of aspect to consider after training probing models is threshold to choose most discriminative features. We pick only these features for which absolute value of corresponding feature is bigger than the threshold. Visualisation of number of features for each threshold can seen in Figure~\ref{fig:appendix_probes_coefs_analysis}.

\end{document}